\documentclass[11pt]{article}
\usepackage{a4,graphicx,xspace,oupbib,amsmath,amsthm}

\newcommand{\be}{\begin{equation}}
\newcommand{\bel}[1]{\begin{equation}\label{#1}}
\newcommand{\ee}{\end{equation}}

\newcommand{\var}{\text{var}}

\newcommand{\C}{\mathcal{C}}
\renewcommand{\S}{\mathcal{S}}

\newcommand{\N}{\mathcal{N}}

\newcommand{\CX}{C_X}
\newcommand{\CY}{C_Y}
\newcommand{\NX}{\mathcal{N}_X}
\newcommand{\NY}{\mathcal{N}_Y}

\newtheorem{proposition}{Proposition}

\begin{document}
\title{{\sc Sampling decomposable graphs using\\ a Markov chain on junction trees}}
\author{
Peter J. Green\thanks {School of Mathematics, University of
Bristol, Bristol BS8 1TW, UK.
\newline \hspace*{5mm} Email: {\tt P.J.Green@bristol.ac.uk}.}\\
University of Bristol.\\
\and 
Alun Thomas\thanks {Division of Genetic Epidemiology, Department of Internal Medicine, University of Utah.
\newline \hspace*{5mm} Email: {\tt Alun.Thomas@utah.edu}}\\
University of Utah.
}
\date{\today}
\maketitle

\begin{abstract}
Full Bayesian computational inference for model determination in undirected graphical models is currently restricted to decomposable graphs or other special cases, except for small scale problems, say up to 15 variables. In this paper we develop new, more efficient methodology for such inference, by making two contributions to the computational geometry of decomposable graphs. The first of these provides sufficient conditions under which it is possible to completely connect two disconnected complete subsets of vertices, or perform the reverse procedure, yet maintain decomposability of the graph. The second is a new Markov chain Monte Carlo sampler for arbitrary positive distributions on decomposable graphs, taking a junction tree representing the graph as its state variable. The resulting methodology is illustrated with numerical experiments on three specific models.

\vspace{5mm}

\noindent {\small {\em Some key words:} 
conditional independence graph,
graphical model,
Markov chain Monte Carlo, 
Markov random field,
model determination.}

\end{abstract}

\section{Graphical modelling and decomposable graphs}
\subsection{Introduction}

Bayesian model determination in the context of graphical models typically involves simultaneous structural and quantitative learning: joint inference about both model parameters $\theta$ and conditional independence graph $G$ in a model for data $Y$ of the form $p(Y\mid \theta,G)$, assuming a joint prior of the form $p(\theta\mid G)p(G)$. Much of the literature considers only the case where $G$ is assumed decomposable, and that is the line taken here, where we develop Markov chain Monte Carlo methods for computational inference in this setting. The assumption of decomposability is a severe restriction, and in our final section we discuss the heavy computational penalty incurred if we relax this assumption.

\textcite{Giudici+Green:99} introduced a reversible jump Markov chain Monte Carlo sampler for posterior sampling of decomposable graphical models, described and implemented for the Gaussian case, which exploited a junction tree representation of decomposable graphs. This allows rapid checking of decomposibility for modified graphs and implementation of modifications, through local computation. The state variable in any such Markov chain must include a representation of the graph, along with associated parameter values. In that previous work, the decomposable graph itself was represented explicitly in the state variable. In this paper, we derive a more efficient sampler that augments the state variable by using a particular junction tree, interpreted as an unknown parameter, not only for the decomposable graph it represents.

We also make an important generalisation applicable to both samplers, allowing certain multiple-edge updates to the graph, maintaining decomposibility, and not only the single-edge moves seen in earlier work. Using this broader class of moves may improve performance in some situations. Our characterisation of a class of multiple-edge perturbations to a graph that maintain decomposability is likely to find broader application in computational graph theory, not only in the Markov chain Monte Carlo sampling of such graphs considered here.
Our paper exploits \textcite{Thomas+Green:09e} which provides an auxiliary result enumerating the number of junction trees corresponding to a given decomposable graph.

As with the graph-updating MCMC sampler introduced in \textcite{Giudici+Green:99}, the new junction tree sampler can be used in conjunction with appropriate parameter-updating methods in a variety of statistical models, and not only to gaussian models as originally presented; equally important are multinomial models for discrete data, under hyper-Dirichlet priors for cell probabilities. For such models, the work of \textcite{Tarantola:04} describes efficient MCMC methodology, which could easily be adapted to use the new sampler presented here.

In statistical science, the use of graphical models in inference is now very well-established. Methodologies in which the graph itself is one of the unknowns in the model are becoming common-place, as inference about the conditional independence properties of models fitted to data is a key part of understanding the stucture of data. As a particular example of the 
applicability and flexibility of the present work, the single edge Markov chain Monte Carlo sampler that we introduce in this paper has already been
used in the {\tt FitGMLD} program described by \textcite{Abel+Thomas:11}.
This program fits a multinomial graphical model to the inter
locus correlations between alleles at proximal genetic markers, a
phenomenon usually referred to as linkage disequilibrium. This is a very practical application, on a large scale: by enforcing
some model restrictions that allow a walking window approach,
this implementation has been used on data representing over
100,000 variables assayed on hundreds of individuals. In contrast to the multinomial set up in this application, the present paper concentrates on gaussian models in examples.

\subsection{Preliminaries on graphical models}

We begin by reviewing some definitions and properties of decomposable
graphs and junction trees; these standard ideas are covered more thoroughly by
\textcite{Lauritzen:96}.

Consider a graph $G=(V,E)$ with vertices $V$ and undirected edges $E$.
A subset of vertices $U \subseteq V$ defines an
induced subgraph of $G$ which contains all the vertices $U$ and
any edges in $E$ that connect vertices in $U$.
A subgraph induced by $U \subseteq V$ is complete if all pairs of
vertices in $U$ are connected in $G$. A clique is a complete subgraph
that is maximal, that is, it is not a subgraph of any other complete subgraph.

A graph $G$ is decomposable if and only if the set of
cliques of $G$ can be ordered as $(C_1,\ldots,C_c)$ so that 
for each $i=2,\ldots, c$
\begin{equation}
\mbox{if} \ \  S_i \ = \ C_i \cap \bigcup_{j=1}^{i-1} C_j \ 
\ \mbox{then} \ \ S_i \subset  C_k \ \ \mbox{for some} \ \ k < i;
\end{equation}
$S_i$ may be empty.

This is called the running intersection property.
Decomposable graphs are also known as triangulated
or chordal graphs; the running intersection property is
equivalent to the requirement that every cycle of length 4 or more in $G$
is chorded.

The sets $S_2, \ldots S_c$ 
are called the separators of the graph.
The set of cliques $\{C_1, \ldots C_c\}$ and the collection of 
separators $\{S_2, \ldots S_c \}$ are uniquely determined
from the structure of $G$, however, there may be many orderings that 
have the running intersection property.
The cliques of $G$ are
distinct sets, but the separators are generally not all distinct. 

The significance of decomposability in statistics and probability stems from the 
existence of the clique--separator factorisation. If a random vector $X$ has a decomposable conditional independence graph $G$, then its distribution factorizes as
\begin{equation}\label{eq:cliqsep}
p(X) = \frac{\prod_{i=1}^c p(X_{C_i})}{\prod_{i=2}^c p(X_{S_i})}.
\end{equation}

The junction graph of a decomposable graph has nodes
$\{C_1, \ldots, C_c\}$ and every pair of nodes is connected.
Each link is associated with the intersection, which may be empty, of the two cliques 
that it connects.

For clarity we will reserve the terms 
vertices and edges for the elements of $G$, and call
those of the junction graph and its subgraphs nodes and links. 

A spanning tree of a graph is a subgraph that includes all of the vertices, and is a tree; let $J$ be any spanning tree of the junction graph. 
It has the junction property if for any two cliques $C$ and $D$ of
$G$, every node on the unique path between $C$ and $D$ in $J$ contains
$C \cap D$. In this case $J$ is said to be a junction tree.

Some authors first partition a graph into its disjoint components
before making a junction tree for each component, combining the result
into a junction forest. 
The above definition, however, will allow us to state
results more simply without having to make special provision for nodes
in separate components. In effect, we have taken a conventional junction forest
and connected it into a tree by adding links between the components. Each
of these new links will be associated with the empty set.
Clearly, this tree has the junction property.
Results for junction forests can easily be recovered from the results
we present below for junction trees.

A junction tree for $G$ will exist if and only if $G$ is decomposable,
and algorithms such as the maximal cardinality search of
\textcite{Tarjan+Yannakakis:84} allows a junction tree representation to
be found in time of order $|V|+|E|$, where $|\cdot|$ denotes the cardinality of a set.
The collection of clique intersections associated with the $c-1$ 
links of any junction
tree of $G$ is equal to the collection of separators of $G$.
The junction property ensures that the subgraph of a junction tree 
induced by the set of cliques that 
contain any set $U \subseteq V$ is a single connected tree.

\subsection{Elaborating the model to include the junction tree}

Each decomposable graph $G$ can be equivalently represented by one or more junction trees. \textcite{Thomas+Green:09e} derived an expression for $\mu(G)$, the number of equivalent junction trees. Given a probability distribution $\pi(G)$ on decomposable graphs, which might for example, be the posterior distribution of the conditional independence graph of a multivariate distribution given data, we can define a distribution on junction trees simply by 
$$
\widetilde{\pi}(J) = \frac{\pi\{G(J)\}}{\mu\{G(J)\}}
$$
where $G(J)$ is the decomposable graph represented by $J$; that is, conditional on $G$ distributed as $\pi(G)$, $J$ is distributed uniformly at random from among the $\mu(G)$ equivalent junction trees.
We assume throughout that $\pi(G)>0$ for all decomposable $G$, so that $\widetilde{\pi}(J)>0$ for all junction trees $J$.

We will construct an ergodic Markov chain whose states are junction trees, with invariant distribution $\widetilde{\pi}$.

\section{Allowable perturbations to decomposable graphs}

\subsection{Single-edge perturbations}

We first follow previous work \cite{Giudici+Green:99} in concentrating on Markov chain Monte Carlo moves that perturb the graph in a very simple way: they connect or disconnect two vertices $x$ and $y$ by adding or removing an edge between them. In general, such a move may destroy the decomposability of the graph, and it is therefore necessary either to test that the perturbed graph is decomposable, or in some way to limit the choice of $(x,y)$ to guarantee in advance that it is decomposable.

\textcite{Frydenberg+L:89} and \textcite{Giudici+Green:99} gave efficient methods for checking that
the perturbed graph
$G'$ is decomposable, given that $G$ is, when the perturbation scheme
involves either connecting or disconnecting an arbitrary pair of vertices.
Using our definition of a junction tree, we can restate their results
as follows.
\begin{itemize}
\item[(C)]
Connecting $x$ and $y$ by adding an edge $(x,y)$ to $G$ will result in a decomposable graph
if and only if $x$ and $y$ are contained in cliques that are adjacent in some junction tree of $G$.
\item[(D)]
Disconnecting $x$ and $y$ by removing an edge $(x,y)$ from $G$ will result in a decomposable graph
if and only if $x$ and $y$ are contained in exactly one clique.
\end{itemize}

\begin{figure}[htbp]
\begin{center}
\resizebox{100mm}{!}{\includegraphics{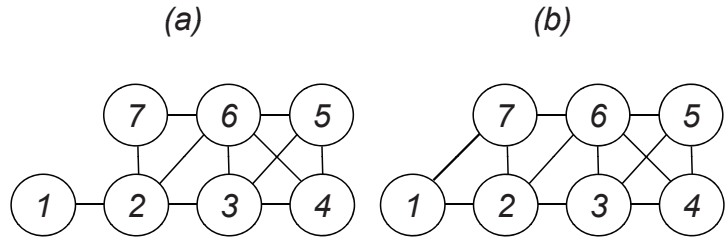}}
\caption{Two small decomposable graphs, differing by the presence of a single edge.}
\label{fig:7vg}
\end{center}
\end{figure}

\begin{figure}[htbp]
\begin{center}
\resizebox{110mm}{!}{\includegraphics{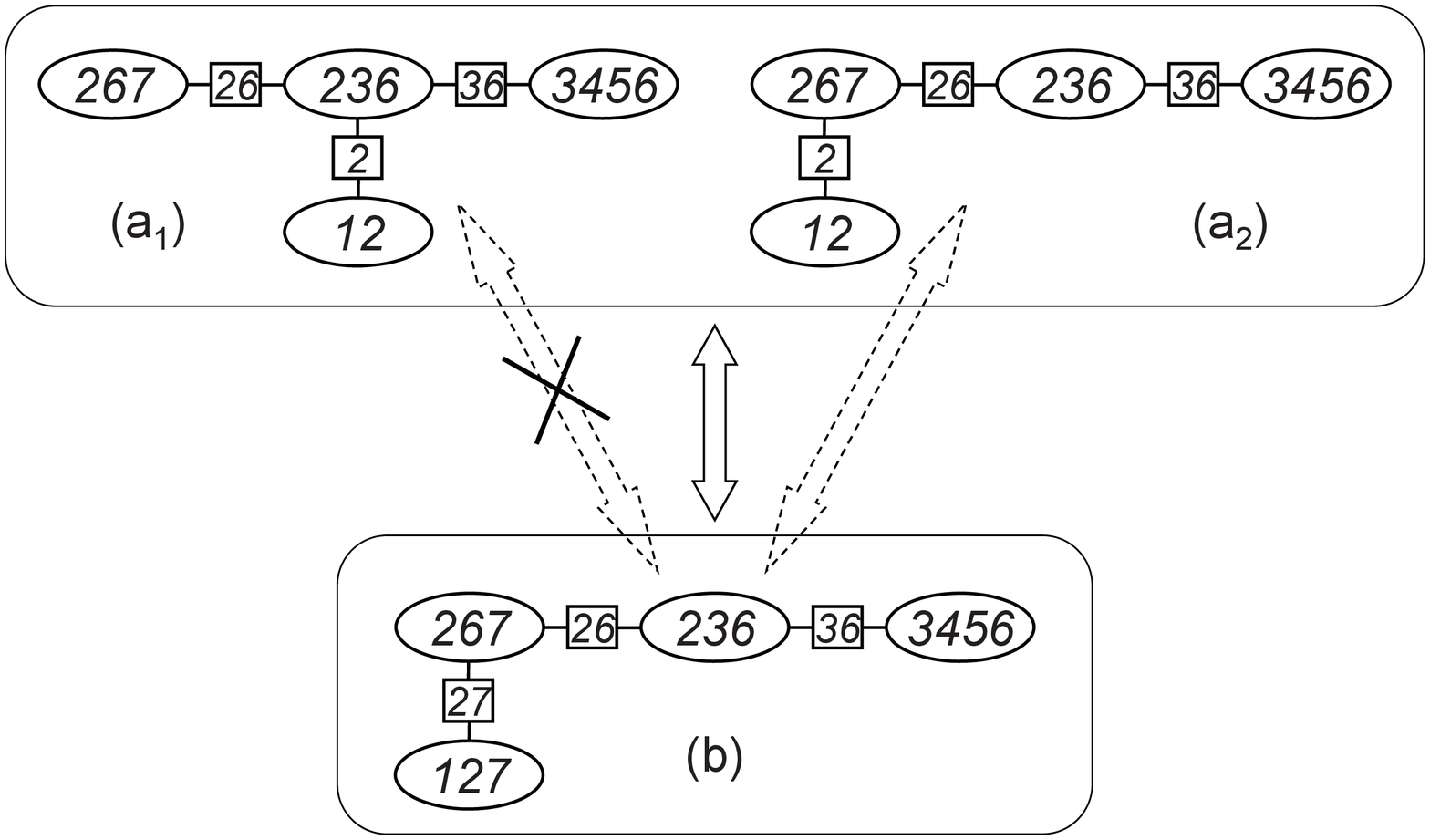}}
\caption{Junction trees corresponding to the decomposable graphs in Figure \ref{fig:7vg}: ellipses represent cliques, and boxes the separators.
Trees (a$_1$) and (a$_2$) correspond to graph (a), and tree (b) to graph (b).}
\label{fig:7vjt2}
\end{center}
\end{figure}

Figure \ref{fig:7vg} illustrates two small decomposable graphs, differing by the presence of a single edge, $(1,7)$. The conditions (C) and (D) are clearly satisfied for this example. Junction trees corresponding to these graphs are shown in Figure \ref{fig:7vjt2}.

This analysis of the single-edge perturbations to a graph that preserve decomposability is key to deriving both the sampler in \textcite{Giudici+Green:99} and that in the present work. In the \textcite{Giudici+Green:99} sampler, the graph-changing move proceeds as follows; it is an example of a reversible jump method \cite{Green:95}. A random pair of distinct vertices is chosen; if they are currently connected, we propose disconnecting them, otherwise we propose connecting them. The algorithm next checks whether the proposed move maintains decomposability, using the results stated above; if it does not, the change is rejected. If it does, then updated parameter values appropriate for the new graph are proposed, and a Metropolis--Hastings acceptance ratio is computed. This leads as usual to acceptance or rejection of the combined graph--parameter update proposal.

The key difference in the new approach lies in replacing the search over junction trees implicit in (C) above with restriction to the current junction tree. In
\textcite{Giudici+Green:99}, where the graph is part of the state variable, we manipulate the junction tree, searching for one for which
the cliques containing $x$ and $y$ are adjacent, and then use that junction tree to effect the perturbation.
In the present work, where the junction tree is part of the state variable, there is no such manipulation, and 
the proposal mechanism is modified so that $x$ and $y$ are only selected
if the cliques containing $x$ and $y$ are already adjacent. 
Figure \ref{fig:7vjt2} illustrates this point. In the sampler of 
\textcite{Giudici+Green:99}, moves between graphs (a) and (b) are possible, even if graph (a) is currently represented by junction tree (a$_1$); the first stage of the move is manipulation from tree (a$_1$) to (a$_2$). However, in the sampler introduced here, moves between trees (a$_2$) and (b) are possible, in either direction, but not between (a$_1$) and (b).

Thus the computational cost savings in our new approach come from the more restrictive
choice of proposed pairs $(x,y)$ specifying edges to be added, and avoidance of the manipulation from one junction tree to another, 
and the price paid is that the space of possible states of the chain is in some sense less connected.
We shall see in Section \ref{sec:expts2} that this price is worth paying, especially in 
larger graphs.

It might be useful at this point to consider for illustration 
the specific but extreme case when
$G$ is the trivial graph with $n$ vertices and no edges. 
The cliques all contain a single vertex, and any tree $J$ connecting these 
vertices
is a valid junction tree, using our generalized formulation. 
$J$ will have $n-1$ edges, and by Cayley's formula \cite{Cayley:89}
we know that it is one of
$n^{n-2}$ possible junction tree representations of $G$.

Under the scheme of \textcite{Giudici+Green:99}, one of the $n(n-1)/2$ possible
pairs of vertices would be selected at random and on inspection and manipulation
of $J$, connecting this pair would be found to make a valid decomposable graph.
With very high probability, $(1-2/n)$, this will require changing $J$ into an 
alternative junction tree $J'$ in which the cliques comprising the selected
vertices of $G$ are connected.

Under our new scheme, one of the $n-1$ links of $J$ would be selected at random.
The vertices making up the cliques that the link joins would be found to contain
a pair of vertices whose connection forms a decomposable graph. The computational
saving is that no manipulation of $J$ is required to establish this. The cost is
that only $n-1$ of the possible $n(n-1)/2$ pairs of vertices can be thus sampled. 
This may be alleviated to some extent by occasionally using the 
randomization step described
by \textcite{Thomas+Green:09e} which allows a junction tree to
be replaced by an equivalent one chosen uniformly at random from the $n^{n-2}$
junction tree representations of $G$.

\begin{figure}[htbp]
\begin{center}
\resizebox{6in}{!}{\includegraphics{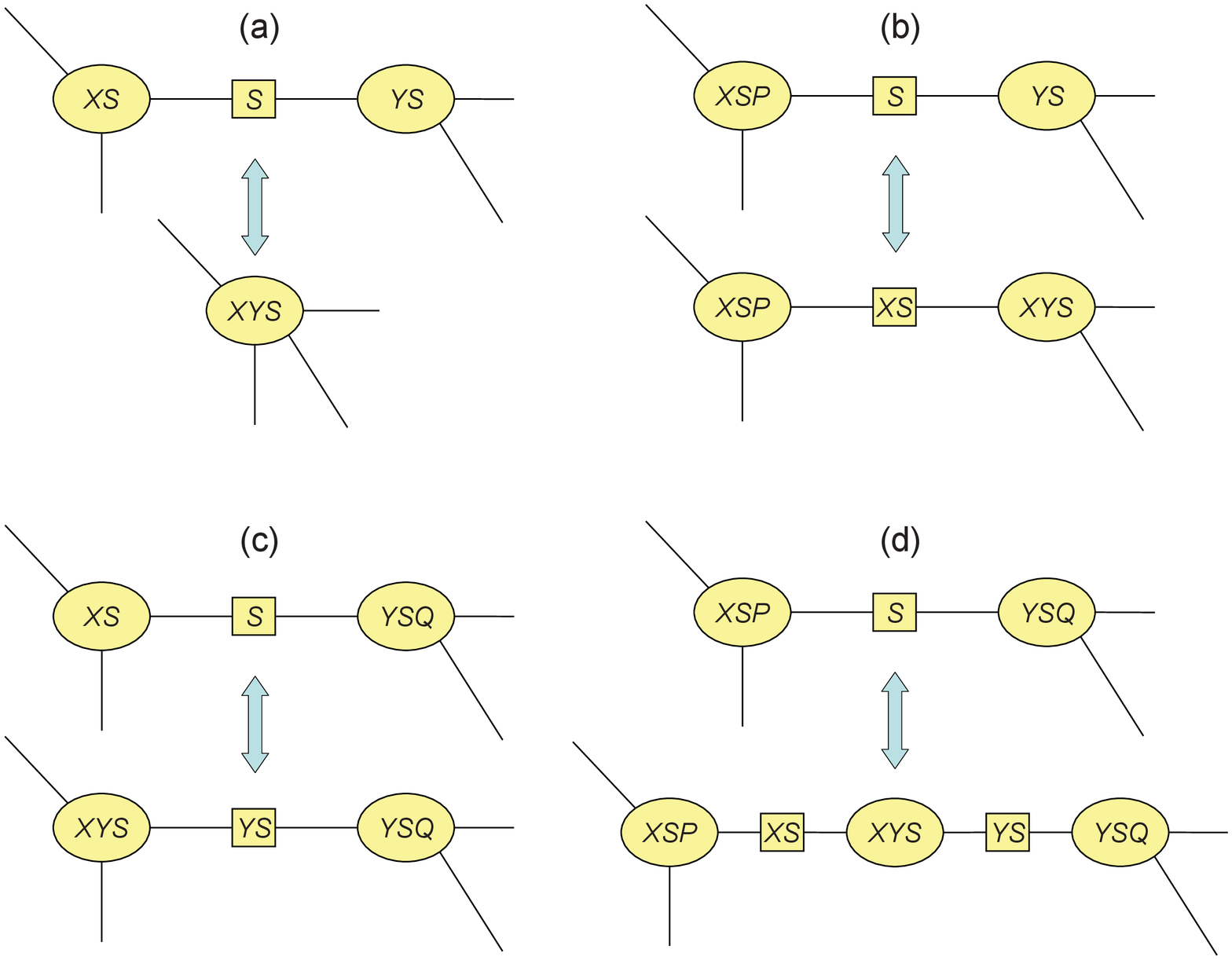}}
\caption{The four possible cases: the clique containing $X=\{x\}$ and $S$ before connecting $X=\{x\}$ and $Y=\{y\}$ is in cases (a) and (c) exactly $XS=X \cup S$, while in cases (b) and (d) it is a proper superset; similarly the clique containing $Y=\{y\}$ and $S$ before the connection is in cases (a) and (b) exactly $YS$ and in (c) and (d) a proper superset. These four cases have to be considered both in the proof that decomposability is maintained, in Section \ref{sec:multi-edge} and Appendix 1, and in the algorithm for making valid connections and disconnections in Section \ref{sec:jtsampler}.}
\label{fig:4cases}
\end{center}
\end{figure}

In both algorithms, the effect on the junction tree of connecting or disconnecting $x$ and $y$ is shown schematically in Figure \ref{fig:4cases}. In each case the upper panel shows part of the junction tree with $x$ and $y$ unconnected;
the lower panel the same part of the tree with them connected. The figures can be read in both directions. The symbol $S$ denotes the separator between the cliques containing $x$ and $y$ referred to in the condition (C) for connecting by adding an edge, and $XYS=\{x,y\}\cup S$ is the clique containing both $x$ and $y$ referred to in the condition (D) for disconnecting by removing an edge. The four cases correspond to the $2 \times 2$ possibilities that the cliques containing $x$ and $y$ are exactly $XS=\{x\}\cup S$ and $YS=\{y\}\cup S$ respectively, or supersets thereof.

These single-edge perturbations to $G$ are a special case of the multiple-edge perturbations defined and justified in the next section, so we omit the proofs that the modifications maintain decomposability. 

\subsection{Multiple-edge perturbations}
\label{sec:multi-edge}

In this section, we present perturbations to decomposable graphs that make multiple connections and disconnections simultaneously, yet are guaranteed to maintain decomposability. Unlike the single-edge moves of the previous section, however, these provide only sufficient, not necessary, conditions for the validity of the perturbations to $G$.

Two disjoint non-empty connected sets of vertices $X$ and $Y$ are said to be completely connected if every vertex in $X$ is connected to every vertex in $Y$. They are completely disconnected if no vertices in $X$ are connected to any vertices in $Y$. 

\begin{proposition}
\label{Cprop}
Suppose $G=(V,E)$ is a decomposable graph, and that $X$ and $Y$ are two disjoint non-empty subsets of $V$ that are each complete in $G$, and which are completely disconnected, i.e. there are no edges $(x,y)$ between any element $x\in X$ and $y \in Y$. Suppose $X$ and $Y$ are subsets of cliques that are adjacent in some junction tree representing $G$.

Let $G'$ be the graph formed from $G$ by completely connecting $X$ and $Y$, i.e. inserting an edge between every pair of vertices $(x,y)$ with $x\in X$ and $y \in Y$.

Then $G'$ is decomposable.
\end{proposition}

\begin{proposition}
\label{Dprop}
Suppose $G=(V,E)$ is a decomposable graph, and that $X$ and $Y$ are two disjoint non-empty subsets of $V$ that are completely connected, i.e., $X\cup Y$ is complete in $G$, such that $X$ and $Y$ are subsets of exactly one clique, $X\cup Y\cup S$, say, where $S\cap (X\cup Y)=\emptyset$. Suppose that one of the following holds:
\begin{enumerate}
\item[(a)] there is no other clique containing $X\cup S$ or $Y\cup S$;
\item[(b)] there is one more clique containing $X\cup S$ but then no other cliques intersecting $X$, and there are no more cliques containing $Y\cup S$;
\item[(c)] there is one more clique containing $Y\cup S$ but then no other cliques intersecting $Y$, and there are no more cliques containing $X\cup S$; or
\item[(d)] there are two more cliques containing $X\cup S$ and $Y\cup S$ respectively, but then no other cliques intersecting $X$ or $Y$, and there is a junction tree $J$ representing $G$ such that there are no other cliques adjacent to $X\cup Y\cup S$ in $J$.
\end{enumerate}

Let $G'$ be the graph formed from $G$ by disconnecting $X$ and $Y$, i.e. removing all edges between pairs of vertices $(x,y)$ with $x\in X$ and $y \in Y$.

Then $G'$ is decomposable.
\end{proposition}

These propositions are presented separately, and it may not be immediately clear that there is a unity to them; in particular it may seem that the conditions in Proposition 2 are much more stringent that those in Proposition 1. In fact, however, they are perfectly matched, since as implemented they precisely delineate the circumstances in which particular moves applied to a junction tree form a reversible pair. 

Thus, in practical use, the junction tree $J$ representing $G$ is already determined before the connection or disconnection of $X$ and $Y$ is considered. Indeed, given $J$ the only $X$ and $Y$ that will ever be considered are those for which this particular junction tree satisfies the conditions mentioned in Proposition 1 and Proposition 2, part (d).  

Finally, we will see that the moves that these propositions confirm maintain decomposability, and can always be implemented by modest local perturbations to the current junction tree.
These local perturbations are illustrated in Figure \ref{fig:4cases}.
\vskip 10pt

The proofs of these propositions are deferred to Appendix 1, following specification in Section \ref{sec:jtsampler} of the algorithms that will implement these perturbations to $G$, which provides further notation and describes the local perturbations of the junction tree in detail. 

Other variant multiple-edge perturbations are possible, but not considered here.

\section{The junction tree sampler}
\label{sec:jtsampler}

\subsection{Multiple-edge connect move}
\label{sec:mc}
The multiple-edge moves of Section \ref{sec:multi-edge} involve choices of appropriate random sets of vertices $X$ and $Y$ in the algorithms detailed below; for the single-edge versions these choices are restricted to be singletons $\{x\}$ and $\{y\}$ respectively, and there are no other changes; therefore, we do not describe the single-edge moves separately.

We first choose a separator $S$ uniformly at random from the collection of 
separators $\S(J)$ in the current junction tree $J$, 
respecting multiplicities of course. 
If $\S(J)$ is empty, which is the case only if the graph consists of a single
clique, no further connection is possible, and we reject immediately.

Suppose $S$ separates cliques $\CX$ and $\CY$: we choose non-empty sets of vertices $X$ and $Y$ from 
$\CX \setminus S$ and $\CY \setminus S$. 
By Proposition 1, completely connecting $X$ and $Y$ yields a new decomposable graph, one junction tree representation of which, $J'$, can be easily formed as follows:
\begin{enumerate}
\item[(a)] if $\CX=X \cup S$ and $\CY=Y\cup S$, then $\CX$, $\CY$ and $S$ are removed from the junction tree, and replaced by a new clique $X \cup Y \cup S$, connected to all those cliques previously connected to $\CX$ or $\CY$, through the same separators as before;
\item[(b)] if $\CX\supset X \cup S$ and $\CY=Y\cup S$, then the vertices in $X$ are added into $S$ and $\CY$, and the junction tree otherwise left unchanged;
\item[(c)] if $\CX=X \cup S$ and $\CY\supset Y\cup S$, then the vertices in $Y$ are added into $S$ and $\CX$, and the junction tree otherwise left unchanged;
\item[(d)] if $\CX\supset X \cup S$ and $\CY\supset Y\cup S$, then the separator $S$ is replaced by a separator / clique / separator triple:
$X \cup S$, $X \cup Y \cup S$, $Y \cup S$, with $\CX$ connected to the first, and $\CY$ to the last, and the junction tree otherwise left unchanged.
\end{enumerate}
These four possibilities are represented graphically in Figure \ref{fig:4cases}, reading downwards.

\subsection{Multiple-edge disconnect move}
\label{sec:md}

For the reverse move, we first draw a clique $C$ at random from the collection
of cliques $\C(J)$ of the current junction tree $J$.
If $C$ contains a single vertex, the proposal is rejected.
We then partition $C$ at random into three sets $X$, $Y$ and $S$, where $X$ and $Y$ at least are non-empty.

The neighbours of $C$ in the junction tree $J$ are then scanned; 
if any neighbour intersects both $X$ and $Y$, 
then disconnecting $X$ and $Y$ is not possible, and the proposal is rejected. 

Otherwise, we partition the neighbours into three sets: 
$\N$, those intersecting neither $X$ nor $Y$, 
$\NX$, those intersecting only $X$, and 
$\NY$, those intersecting only $Y$. 
Among the cliques in $\NX$, we select an arbitrary one of any 
encountered that contains all of $X \cup S$ and identify 
this as $\CX$; if none are encountered, the set is left undefined. Similarly, we 
look in $\NY$ to try to identify $\CY$.

\begin{enumerate}
\item[(a)] If neither of $\CX$ and $\CY$ are defined, then $X$ and $Y$ are disconnectible: we create new cliques $C\setminus Y=X \cup S$ and $C\setminus Y = Y \cup S$, with a separator $S$ between them. The first of these is connected to those cliques in $\NX$ and the second to those in $\NY$. Those in $\N$ 
are connected at random to one
of the new cliques. Finally the clique $C$ is deleted.
\item[(b)] If $\CX$ is defined, but not $\CY$, then disconnection is possible if and only if $\NX$ contains exactly one clique, $\CX$ itself: in this case, $X$ is removed from the clique $C$ and from the adjacent separator $C\setminus Y=X \cup S$ connecting it to $\CX$; the junction tree is otherwise unchanged.
\item[(c)] If $\CY$ is defined, but not $\CX$, then disconnection is possible if and only if $\NY$ contains exactly one clique, $\CY$ itself: in this case, $Y$ is removed from the clique $C$ and from the adjacent separator $C\setminus X=Y \cup S$ connecting it to $\CY$; the junction tree is otherwise unchanged.
\item[(d)] If both of $\CX$ and $\CY$ are defined, then $X$ and $Y$ can only be disconnectible if $\N$ is empty, and both $\NX$ and $\NY$ contain exactly one clique. In this case, the clique $C$ and its adjacent separators $C\setminus Y=X \cup S$ and $C\setminus X=Y \cup S$ are removed from the junction tree, and replaced by a separator $S$ linking the cliques $\CX$ and $\CY$.
\end{enumerate}
These four possibilities are represented graphically in Figure \ref{fig:4cases}, reading upwards.

\subsection{Choices of $X$ and $Y$, and associated proposal probabilities}

Whether using single-edge or multiple-edge updates, in each of the connect and disconnect moves at one point we have to choose sets of vertices $X$ and $Y$ at random, subject to the stated constraints. Providing that the probabilities with which these choices are made are correctly encoded into the Metropolis--Hastings acceptance calculation through the proposal probabilities $q(J,J')$, the junction tree sampler satisfies detailed balance whatever probability distribution for $X$ and $Y$ is used. Varying this choice allows scope for improving performance, although we have not conducted any systematic experiments on this issue.

In the single-edge case, $X=\{x\}$ and $Y=\{y\}$ are both singletons, and we have few options. For the connect move, we choose $x$ and $y$ uniformly at random from $\CX\setminus S$ and $\CY\setminus S$ respectively. The probability $q(J,J')$ that starting from $J$ leads to the proposed modified junction tree $J'$ specified in Section \ref{sec:mc}, following the uniform random choice of $S$, is easily seen to be 
$1/\{|\S(J)|\times (m_X-s) \times (m_Y-s) \}$, 
where $m_X=|\CX|$, $m_Y=|\CY|$ and $s=|S|$.

For the disconnect move, we choose $x$ and $y$ uniformly at random without replacement from $C$; then the process in Section \ref{sec:md} yields the proposal probability $(1/|\C(J)| )  \times \{ 2/m(m-1) \} \times 2^{-|\N|}$
in case (a), and otherwise $( 1/|\C(J)| )  \times \{ 2/m(m-1) \} $, where $m=|C|$. The factor 2 in the numerator accounts for the fact that the effect of the move on the junction tree is not affected by the order in which $X$ and $Y$ are drawn.

Turning to the multiple-edge case, out of wider ranges of options we choose the simplest. 
For the connect move, to select $X$ from $\CX\setminus S$, we first pick $N_X$ uniformly at random between
$1$ and $|\CX\setminus S|$ and then choose $X$ to be a subset of $\CX\setminus S$ of that size chosen
uniformly at random from all such.  We choose $N_Y$ and $Y$ similarly, and independently. The proposal probability is 
$$
\frac{1}{|\S(J)|}\times \frac{1}{m_X-s} \frac{N_X!(m_X-s-N_X)!}{(m_X-s)!}\times \frac{1}{m_Y-s} \frac{N_Y!(m_Y-s-N_Y)!}{(m_Y-s)!}
$$

For the disconnect move, we choose $M$ uniformly at random between $2$ and $m=|C|$,
then $N$ uniformly at random between $1$ and $M-1$. 
We then partition $C$ into sets $X$, $Y$ and $S$ of sizes $N$, $M-N$ and $m-M$, respectively,
uniformly at random from all such partitions. This sampling can be conducted efficiently in a
single pass through $C$. The proposal probability is 
$$
\frac{1}{|\C(J)|} \times \frac{2}{(m-1)(M-1)} \times \frac{N! (M-N)! (m-M)!}{m!}.
$$
Again this has to be multiplied by an additional factor $2^{-|\N|}$ in case (a).

\subsection{Detailed balance, irreducibility and ergodicity}
\label{sec:accept}

The standard Metropolis--Hastings acceptance probability for this proposal \cite{Hastings:70}
is
$$
\alpha(J,J') = \min\left\{1, \frac{\widetilde{\pi}(J')q(J',J)}{\widetilde{\pi}(J)q(J,J')} \right\}
$$
which ensures detailed balance with respect to the target distribution $\widetilde{\pi}(J)$. 

A fact that is well known but not commonly exploited is that the acceptance probability 
expression cited above is not the only choice yielding detailed balance. In particular, consider the alternative
choice of acceptance probability
$$
\widetilde{\alpha}(J,J') = 
\min\left\{1, \frac{\widetilde{\pi}(J')}{\widetilde{\pi}(J)}\right\}
\times
\min\left\{1, \frac{q(J',J)}{q(J,J')} \right\}.
$$
Then the equilibrium joint probability of the chain being in state $J$ followed by $J'\neq J$ is
$$
\widetilde{\pi}(J) q(J,J') \widetilde{\alpha}(J,J') = 
\min\left\{\widetilde{\pi}(J), \widetilde{\pi}(J')\right\}
\times
\min\left\{q(J,J'), q(J',J) \right\},
$$
an expression evidently symmetric in $J$ and $J'$. Thus this chain is also reversible,
with the same invariant distribution $\widetilde{\pi}(J)$. 

According to the important result of \textcite{Peskun:73}, since $\widetilde{\alpha}(J,J') \leq \alpha(J,J')$
for all $J\neq J'$, this new chain is inferior to the Metropolis--Hastings one, in respect of the asymptotic variance
of any ergodic average. However, in computational terms, it may still be advantageous. An accept
decision taken with probability $\widetilde{\alpha}(J,J')$ will involve computing the ratios
$\widetilde{\pi}(J')/\widetilde{\pi}(J)$ and $q(J',J)/q(J,J')$ separately, and comparing with two independent
uniform random numbers. The proposal is rejected if either test fails. Thus in situations where either
of these probability ratios is costly to compute, there is scope for saving time by 
delaying computing the more expensive of the two ratios, only doing so if the pre-test 
using the first ratio is passed.

Turning now to irreducibility, recall that $\widetilde{\pi}(J)>0$ for all $J$. Since the moves of our chain are reversible, it is sufficient to show that there is a path of junction trees, formed by successively adding edges one by one, from any $J$ up to the trivial junction tree, with all vertices in a single clique, corresponding to the completely connected graph. But we can always add an edge to any junction tree other than this trivial one, simply by selecting a pair of disconnected vertices in adjacent cliques, and connecting them.

The state space of the chain is finite, so it follows from this irreducibility that the chain is ergodic, and so ergodic averages converge to expectations under the invariant distibution $\widetilde{\pi}$.

\section{Numerical experiments and performance of the new sampler}
\label{sec:expts}
\subsection{Introduction}

We present three numerical illustrations of the new sampler in operation. First,
focussing solely on testing the sampler on the graph model alone, in the absence of parameters or data, we 
show that for decomposable graphs on $n=7$ vertices we can correctly sample either 
uniformly over junction trees or uniformly over decomposable graphs. For the 
second illustration we introduce a novel graphical Gaussian intra-class model 
from which we simulate data and then use our approach to sample from the posterior 
distribution of models given the simulated data. Finally we give an example where we use an annealing approach to infer the decomposable graph most strongly supported by the data.

The programs to carry out these computations were written in Java and are included
in the {\em Java Programs for Statistical Genetics and Computational Statistics}
package that can be obtained from
http:/balance.med.utah.edu/wiki/index.php/JPSGCS.

\subsection{Decomposable graphs of size 7}

Using a brute force approach we iterated through all 2,097,152 undirected 
graphs on seven labelled vertices and identified the 617,675 decomposable ones.
A list of the cliques of each decomposable graph was found and used as
an index into a table of counters.
The number of possible junction tree representations for each graph was found 
using the algorithm given by \textcite{Thomas+Green:09e} and recorded. The
decomposable graphs were sorted from those with most representations to least; there are 16,807 junction trees for the trivial graph, and 187,447 graphs with a single junction tree.

We began with $G$ set as the trivial graph and chose $J$ uniformly at random from
the possible representations. 
For each simulated junction tree, the list of cliques comprising its nodes were
used to find the appropriate counter in the indexed table, which was updated. 

In the first case we sampled uniformly over junction trees, that is
with $\widetilde{\pi}(J) \propto 1$, and, hence, $\pi\{G(J)\} \propto \mu\{G(J)\}$. In the
second case we set $\widetilde{\pi}(J) \propto 1/\mu\{G(J)\}$ which should give
a uniform sample of decomposable graphs. The value of $\mu\{G(J)\}$ is directly computable
from $J$ and does not require the construction of $G(J)$.
In each case we sampled 1,000,000 graphs. The times taken for the runs were
70 and 76 seconds respectively, but the first 60 seconds in each case
was used to make the indexed
table, a step not typically required in a real application.

\begin{figure}[htbp]
\begin{center}
\resizebox{100mm}{!}{\includegraphics{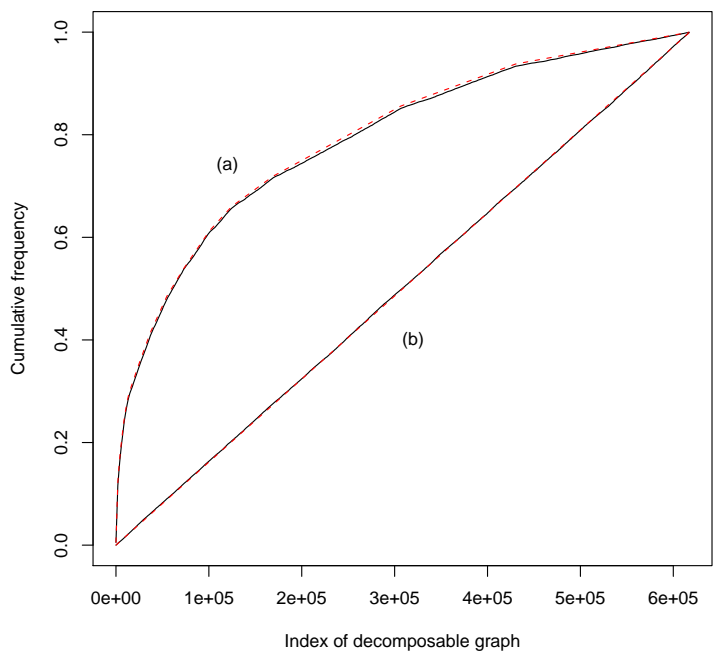}}
\caption
{
Cumulative distribution functions for decomposable graphs of size seven sampled
(a) with probability proportional to the number of junction tree representations
and (b) uniformly. The solid lines give the observed frequencies and the expected
distributions are shown by the dashed lines. The graphs are indexed from left
to right in decreasing order by number of junction tree representations.
}
\label{fig.seven}
\end{center}
\end{figure}

Figure \ref{fig.seven} compares the expected and empirical distribution 
functions for both of these runs, and shows an excellent correspondence.
Similar performance was observed for both standard Metropolis--Hastings sampling and the 
variant described above. 

\subsection{A graphical Gaussian intra-class model}
\label{sec:expts2}
Given a decomposable graph $G$ on $v$ vertices labelled $1,\ldots,v$, and real scalar parameters $\sigma^2>0$ and $\rho$, we define a non-negative definite matrix $\Sigma=\Sigma_G(\sigma^2,\rho)$ by
$$
\Sigma_{ij} = \begin{cases} 
\sigma^2, &  i=j \\
\rho\sigma^2, & (i,j) \text{ is an edge in } G,
\end{cases}
$$
and $(\Sigma^{-1})_{ij}=0$ if $(i,j)$ is not an edge in $G$. 

By \textcite{Grone:84}, since $G$ is decomposable and $\Sigma$ restricted to each clique is positive definite, $\Sigma$ exists and is unique, in fact the unique completion of the specified entries that is positive definite; it is the variance matrix of a $v$-variate Gaussian distribution for which $G$ is the conditional independence graph. We call this the graphical Gaussian intra-class model.

Suppose that $y\sim N\{0,\Sigma_G(\sigma^2,\rho)\}$. Then if $\C$ and $\S$ denote the sets of cliques and separators of $G$, the clique-separator factorisation (\ref{eq:cliqsep}) gives
\bel{eq:cliqsepgauss}
p(y\mid G,\sigma^2,\rho) = \frac{\prod_{C\in\C} p(y_C\mid G,\sigma^2,\rho)}{\prod_{S\in\S} p(y_S\mid G,\sigma^2,\rho)}.
\ee
Since each $C$ is a complete subgraph of $G$, $\var(y_C)$ is explicitly specified in the assumptions above, it is the intra-class model
$\sigma^2 \{(1-\rho)I_C +\rho J_C\}$ where $I_C$ and $J_C$ are respectively the identity matrix and the matrix of all ones, with rows and columns both indexed by $C$. But the inverse and determinant of this variance matrix may be written down explicitly, and so we have
\begin{multline}\label{eq:ggimcliq}
p(y_C\mid G,\sigma^2,\rho) = (2\pi)^{-v_C/2} \sigma^{-v_C} \left\{(1-\rho)^{v_C-1} (1-\rho+v_C\rho)\right\}^{-1/2}
\times \\ 
\exp\left\{\frac{-1}{2\sigma^2(1-\rho)}(y_C^Ty_C-\frac{\rho}{1-\rho+v_C\rho}y_C^TJ_Cy_C)\right\}
\end{multline}
where $v_C$ is the number of vertices in $C$. Replacing $C$ by $S$ throughout, the same holds for each $p(y_S\mid G,\sigma^2,\rho)$. Noting that $\sum_{C\in\C}v_C-\sum_{S\in\S}v_S = v$, we thus have the joint distribution
explicitly, from (\ref{eq:cliqsepgauss}):
\begin{multline*}
p(y\mid G,\sigma^2,\rho)=(2\pi)^{-v/2} \sigma^{-v} (1-\rho)^{-v/2} \prod_{C\in\C} \{1+v_C\rho/(1-\rho)\}^{-1/2}
\prod_{S\in\S} \{1+v_S\rho/(1-\rho)\}^{+1/2}
\times \\
\exp\left[\frac{-1}{2\sigma^2(1-\rho)}
\left\{\sum_{C\in\C}\left(y_C^Ty_C-\frac{\rho}{1-\rho+v_C\rho}y_C^TJ_Cy_C\right)
-\sum_{S\in\S}\left(y_S^Ty_S-\frac{\rho}{1-\rho+v_S\rho}y_S^TJ_Sy_S\right)
\right\}
\right],
\end{multline*}
which can be simplified to
\begin{multline}\label{eq:joint}
p(y\mid G,\sigma^2,\rho)=(2\pi)^{-v/2} \sigma^{-v} (1-\rho)^{-v/2} \prod_{C\in\C} f(C)^{-1/2}
\prod_{S\in\S} f(S)^{+1/2}
\times \\
\exp\left[\frac{-1}{2\sigma^2(1-\rho)}
\left\{y^Ty -\rho \sum_{C\in\C} H(C)
+\rho \sum_{S\in\S} H(S)
\right\}
\right],
\end{multline}
where $f(D)=\{1+v_D\rho/(1-\rho)\}$ and $H(D)=(\sum_{i\in D} y_i)^2/(1-\rho+v_D\rho)$ for any $D\subseteq \{1,\ldots,v\}$.

The necessary and sufficient condition on $\rho$ for this distribution to be well-defined for all decomposable graphs $G$ on $v$ vertices is that $-1/(v-1)<\rho<1$.

\begin{figure}[htbp]
\begin{center}
\resizebox{100mm}{!}{\includegraphics{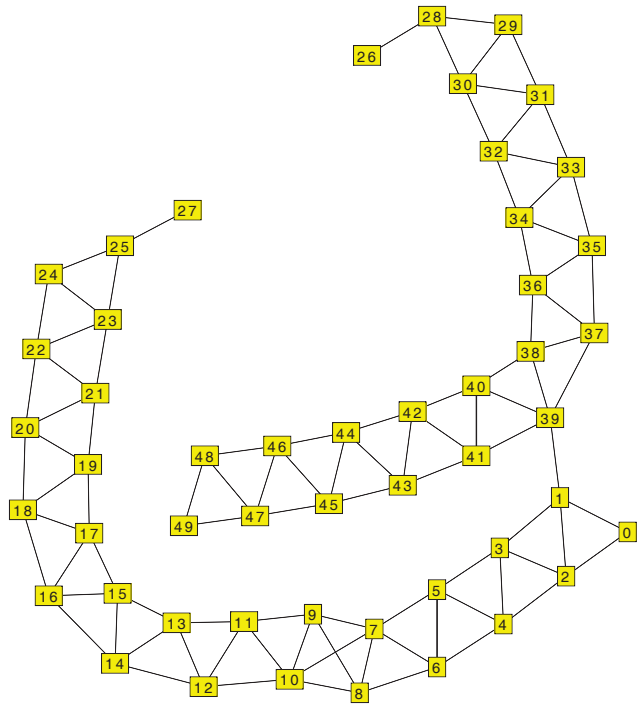}}
\caption
{
A graph typical of the type sampled early in their runs by all three samplers for the graphical Gaussian intra-class model of Section \ref{sec:expts2}, on $v=50$ vertices.
The edge between variables 1 and 39 is spurious, and has to be removed
before the correct edges near variables 25 and 26 can be added.
}
\label{fig.graph}
\end{center}
\end{figure}

Using the method in Appendix 2, we simulated 1000 graphical Gaussian intra-class model observations 
on $50$ variables with $\sigma^2 = 30$ and $\rho = 0.2$. We used a second order
Markov chain graphical structure, that is, $(\Sigma^{-1})_{ij} = 0$ for all $i$ and 
$j$ such that $|i-j| > 2$. This data set is denoted by $D$ below.

For a Bayesian analysis of these synthetic data, we assume the graphical Gaussian intra-class model, and place independent priors on $\sigma^2$, $\rho$ and $G$: 
$\sigma^{-2}\sim \text{Gamma}(\alpha,\beta)$, $\rho$ is uniform on the valid range $(-1/(v-1),1)$
and $G$ is uniform on all decomposable graphs.
It is clear from (\ref{eq:joint}) that the inverse Gamma distribution is conditionally conjugate for $\sigma^2$ in this model, so the posterior full conditional for $\sigma^{-2}$ is
$$
\sigma^{-2}|\rho,G,y \sim \text{Gamma}(\alpha+nv/2,\beta+Q/\{2(1-\rho)\}),
$$
where $Q=\sum_{r=1}^n (y^{(r)})^Ty^{(r)} 
-\rho \sum_{C\in\C} H(C)+\rho \sum_{S\in\S} H(S)$. 
Thus there is a straightforward Gibbs sampler update for $\sigma^2$. 

On the other hand, we must use a Metropolis--Hastings update for $\rho$, as the full conditional is non-standard. In view of the constraint on $\rho$, we suggest a symmetric additive random-walk Metropolis proposal on the logistic-like transform $g(\rho)=\log[\{\rho+1/(v-1)\}/(1-\rho)]$. Thus we set $\rho^\star=g^{-1}\{g(\rho)+z\}=1-\{v/(v-1)\}/[\exp\{g(\rho)+z\}+1]
=1-\{v/(v-1)\}/(e^z[\{\rho+1/(v-1)\}/(1-\rho)]+1)$, where the innovation $z$ has any distribution symmetric about 0.

The acceptance probability for detailed balance with respect to the posterior distribution will be
$$
\alpha=\min\left\{1,\frac{p(\rho^\star)p(y\mid G,\sigma^2,\rho^\star)g'(\rho)}{p(\rho)p(y\mid G,\sigma^2,\rho)g'(\rho^\star)}\right\}.
$$
Since $g'(\rho)=\{v/(v-1)\}/[\{\rho+1/(v-1)\}(1-\rho)]$, this becomes
$$
\alpha=\min\left\{1,\frac{p(\rho^\star)p(y\mid G,\sigma^2,\rho^\star)\{\rho^\star+1/(v-1)\}(1-\rho^\star)}{p(\rho)p(y\mid G,\sigma^2,\rho)\{\rho+1/(v-1)\}(1-\rho)}\right\}.
$$
The distribution $p(y\mid G,\sigma^2,\rho)$ is given in (\ref{eq:joint}), and while there is some cancellation, this is still quite a cumbersome calculation. 

For the graph, we use one of three different Metropolis--Hastings update moves: a junction tree sampler that 
proposes single edge connections or deletions, a junction tree sampler that
proposes multiple edge updates, and the Giudici--Green sampler.
The first and second use the theory of Section \ref{sec:jtsampler}, with target distribution $\widetilde{\pi}(J)$ replaced by $\widetilde{\pi}(J\mid Y,\sigma^2,\rho)$. 

Using these parameter update moves, we sampled from the joint posterior distribution 
of $G$, $\sigma^{2}$ and $\rho$.
In each case we started from the initial conditions
of $\sigma^{2} = 1$, $\rho = 0$ and $G$ set to have no edges indicating complete
independence between the $50$ variables.
We made 1,000,000 Metropolis--Hastings updates with each sampler and 
output
values indicating the state of the chain after every 100 iterations.
The
parameters $\sigma^{2}$ and $\rho$ were updated as described above after each 
1,000 Metropolis--Hastings steps. For the junction tree samplers we also 
randomized the junction tree after every 1,000 Metropolis--Hastings steps using
the method given by \textcite{Thomas+Green:09e}. 

The computations of log likelihoods under the graphs $G$ decompose
into sums of contributions, or scores, from the subsets of vertices that
are the cliques and separators of $G$. Using the ideas in Appendix 3, there are significant cancellations in the log likelihood ratios used when updating.
The score associated with a subset of vertices depends on 
$\sigma^{2}$, $\rho$ and the appropriate sufficient statistics, but not 
on $G$. Hence, in our implementation,
after computing the score of a subset, its sufficient statistics are cached and
indexed by the elements of the subset. This avoids recomputation and in the long
run makes the running time of
our samplers independent of the number of observations in the sample.

Visual inspection of trace plots of the log likelihood of sampled states,
and the sampled values of $\sigma^{2}$ and $\rho$ (not shown) reveal similar sampling
properties for the three samplers, on a sweep-by-sweep basis. The posterior distributions for $\sigma^2$ and $\rho$ are centred close to the true values.
The variance stabilizes almost
immediately while the correlation takes longer, around 100,000 sweeps, to converge, requiring that the current graph estimate is close to correct first. 

We also monitored acceptance rates and run times.  
The former varied between runs, but there is a consistent pattern
of the single edge junction tree sampler accepting more
proposals than the multi edge junction tree sampler, which in turn accepts more 
than the Giudici--Green sampler.
The running times were very consistent between runs: the Giudici--Green sampler took about 70 seconds for the 1,000,000 sweeps, while the two junction tree samplers both took about 8 seconds. The greater running time
for the Giudici--Green sampler is due to the necessity of searching and updating
the junction tree to find proposals that result in decomposable graphs. This
outweighs the time required by the junction tree methods to compute
$\mu(J)$ and to perform the junction tree randomization steps.
The randomization step was found to be necessary with poor graph reconstructions
when it was omitted. However, its omission did not greatly affect
estimation of $\sigma^{2}$ and $\rho$.

\begin{figure}[htbp]
\begin{center}
\resizebox{140mm}{!}{\includegraphics{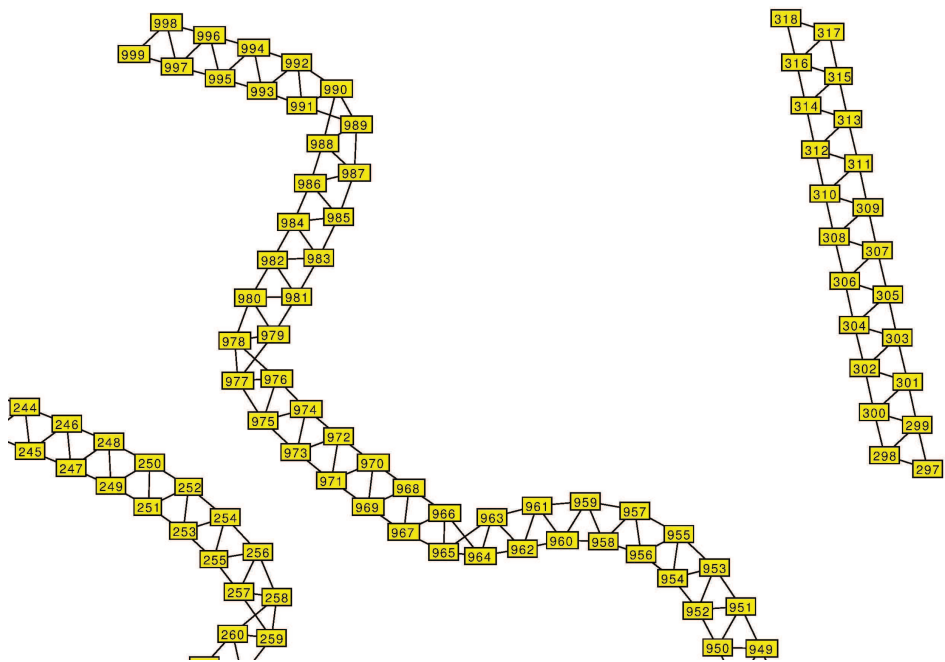}}
\caption
{
Similar to Figure \ref{fig.graph}, but for $v=1000$ vertices, zoomed in to show detail.
}
\label{fig.graph2}
\end{center}
\end{figure}
Figure \ref{fig.graph} shows an inappropriate graph typical of those that
all of the samplers visit early in the simulation. There is, in the 
data, a strong
but in fact spurious correlation between variables 1 and 39, and the corresponding
edge appears in this graph. Because only decomposable graphs are sampled, 
the presence of this edge prevents the correct edges elsewhere in the graph 
from being formed; adding the correct edges would make a long
loop of the type that is prohibited in decomposable graphs. For the
Giudici--Green sampler, getting to the more probable states requires a sequence
of steps that first removes the edge between 1 and 39 and then adds one
between 25 and 26, or similar. The sequence of moves required by the junction
tree samplers is more complex requiring the deletion of edge 1 to 39, then
the randomization of the junction tree to give, for example, one that has the clique
$\{25,27\}$ adjacent to $\{26,28\}$, and then the connection of 25 to 26, or similar.
Despite the extra requirement of an appropriate junction tree configuration,
all the samplers eventually make the transition into the appropriate 
part of the graph space. 
Although not shown here, the most probable graph, as sampled by all three methods,
was similar to the one used to generate the data, but was missing the 
edge between 24 and 26, which we attribute to simple sampling variation.

Also seen in Figure \ref{fig.graph} is an edge between variables 
7 and 10 that was not in the generating model. Small local changes such as 
this appear and vanish throughout the sampling run.

This example may be run successfully on a much larger scale. Figure \ref{fig.graph2} shows a zoomed fragment of a posterior sample derived from a graphical Gaussian intra-class model on $v=1000$ vertices. As can be seen, the inferred graph is not connected: there are in fact 12 components in the whole graph at this stage of the simulation. However, there are no false positives; every edge in this inferred graph is an edge in the true graph. The set up for this example is identical to that for the 50-vertex example above, except that $n=2000$ data vectors were generated from the true graphical Gaussian intra-class model, a graph prior of the form $p(G)\propto \exp(-2 |E|)$ on decomposable graphs was used, where $|E|$ is the number of edges in $G=(V,E)$, and the junction tree sampler was run for 80,000,000 sweeps, taking about 5 hours.

\subsection{Fitting a decomposable model to synthetic data}
\label{sec:joneseg}
For a final example, we revisit the 15-vertex decomposable model considered in Section 7 of \textcite{Jones:05}, using their synthetic Gaussian data set.
We use the junction tree and Giudici--Green samplers to estimate the best-supported decomposable graph.
Rather than conduct a Bayesian analysis, for this example we fixed the variances and covariances at their maximum likelihood estimates for each decomposable graph $G$, and considered the maximum penalized profile likelihood under the assumed model, using an additive penalty on the graph $G$ equal to $-\alpha|E|$ on the log-likelihood scale, where $\alpha=\log(\{|V|-1\}/d-1)$ and $d=1$ or $2$. This penalty corresponds to the graph prior used by Jones et al., in which edges are generated as Bernoulli trials; our parameter $d$ is the expected average vertex degree, and they state that their experiments use $d=2$.
We adopt a simulated annealing approach to estimating the maximum penalized profile likelihood decomposable graph, using a geometric cooling schedule with factor 0.999999. 

\begin{figure}[htbp]
\begin{center}
\resizebox{60mm}{!}{\includegraphics{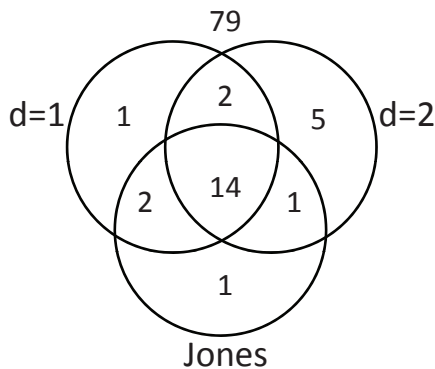}}
\caption
{
Analysis of all possible $\binom{15}{2}=105$ edges according to their presence in the optimum graph in the analysis of Section \ref{sec:joneseg}, with $d=1$ and $2$, and the highest posterior probability graph of \textcite{Jones:05}.
}
\label{fig.venn}
\end{center}
\end{figure}

In our experiments, we made 250 independent replications of runs of length 3,000,000 sweeps of each of the single-edge and multiple-edge junction tree samplers, and the Giudici--Green sampler
for $d=1$: in each case, at least 98\% of these runs visited the optimum graph, i.e. that with highest penalized likelihood, and in at least 17\% of the runs, the optimum graph was the final state of the simulation.  In case $d=2$, these minimum proportions were 96\% and 16\%. These runs took about 13 seconds each. Figure \ref{fig.venn} summarizes the correspondence between our optimum graphs and the highest posterior probability graph of \textcite{Jones:05}, as presented in their Figure 5. 

Our optimum inferred graphs attain higher posterior probability than that reported by \textcite{Jones:05}, in both cases $d=1$ and $2$, although that is their objective. In separate, longer, annealing runs, we estimated the posterior probabilities relative to the optimum; in Table \ref{tab.modes} we report these relative probabilities both as calculated algebraically, and on the basis of relative frequencies.

\begin{table}[htbp]
\begin{center}
\caption
{
Posterior probabilities of selected inferred graphs, based on calculation and on empirical frequencies.
}
\label{tab.modes}
\vspace{3mm}
\begin{tabular}{clcc}
\hline
analysis & graph & calculated & sampled \\
\hline
& Jones highest probability & 0.00742 & 0.00566 \\
$d=1$ & Optimum, $d=1$ & 1 & 1 \\
& Optimum, $d=2$ & 0.0698 & 0.0766 \\
\hline
& Jones highest probability & 0.00464 & 0 \\
$d=2$ & Optimum, $d=1$ & 0.3001 & 0.0217 \\
& Optimum, $d=2$ & 1 & 1 \\
\hline
\end{tabular}
\end{center}
\end{table}

\section{Discussion}

\subsection{Other recent work on decomposable Gaussian graphical models}

Although the restriction to decomposable graphs is a substantial one, it has been a common assumption in much of recent methodological research on graphical models. 

For Gaussian decomposable models, recent developments include \textcite{RajaratnamMC:08}, which has an extensive development of methodology for estimating $\Sigma$ for a fixed known decomposable graph $G$, using a rich class of priors $p(\Sigma|G)$. It shows that a decision theoretic approach can outperform simple MLEs. There is some discussion of estimating $G$, but only in the context of an example where the number of possible $G$ is 60, small enough for the marginal posterior probability for every possible graph can be computed and compared. This complete enumeration approach is clearly not a feasible one in the general case. There may be scope for combining this approach to parameter inference with our method for sampling over graphs.

\textcite{CarvalhoS:09} discuss the impact of the choice of priors on inference about $G$ when fitting decomposable graphical models. There is some comment on the choice of $p(G)$, but the paper concentrates on the specification of $p(\Sigma | G)$ which these authors contend should be a hyper-inverse Wishart g-prior. Again, these ideas might be used in conjunction with our new sampler.

\textcite{ScottC:08} describes an optimization method called Feature-Inclusion Stochastic Search, for fitting decomposable Gaussian graphical models. Note that it does not consider non-decomposable models for much the same reasons that we do not in detail, although this fact is not mentioned in the title or abstract, and it aims to find high-probability graphs but not to approximate the whole posterior distribution. Unfortunately, their discussion of the \textcite{Giudici+Green:99} sampler is inaccurate in saying that it is limited to problems of up to about 6 vertices, and in asserting that MCMC methods require expensive complete enumerations of all possible updates that yield decomposable proposals.

\subsection{Other recent work on Gaussian graphical models under different restrictions}

Two recent papers have discussed the graphical model choice problem for particular restricted classes of graphs. They both rely on results from \textcite{Atay-Kayis:05}, and so apply only to Gaussian models.

\textcite{LenkoskiD:11} describe the Mode Oriented Stochastic Search algorithm, an extension of a random uphill search that maintains a list of the $m$ most probable graphs visited during a run of the search. The approach is not restricted to decomposable graphs, but it does restrict the maximum clique size in the graphs considered (to 5 or less in their example). This restriction cannot be relaxed, since stages of the algorithm involve some very expensive computations. For instance, the first step of the ``iterative proportional scaling algorithm and the block Gibbs sampler" requires that you find all the cliques of an arbitrary undirected graph. This is an NP hard problem. The algorithm is focussed on (approximate) optimization, and cannot be used for sampling from posterior distributions on graphs.

\textcite{DobraLR:11} discuss model estimation for Gaussian models where the graphs are restricted to subsets of multivariate lattices, and this methodology is therefore limited to particular classes of spatial statistics problems.

\subsection{Other recent work on Gaussian graphical models that are not necessarily decomposable}

If $G$ is not decomposable, (\ref{eq:cliqsep}) does not hold, but we still have the prime component factorisation
$$
p(X) = \frac{\prod_{i=1}^c p(X_{P_i})}{\prod_{i=2}^c p(X_{S_i})}
$$
where the prime components $P_i$ are the maximal subgraphs that cannot be decomposed: in a non-decomposable graph, at least one is not complete.

None of the analysis works as cleanly as for decomposable graphs, but broadly analogous 
model formulations and sampling methods have been discussed by, for example,
\textcite{Dellaportas:99},
\textcite{Roverato:02},
\textcite{Dellaportas:03} and
\textcite{Atay-Kayis:05}.

The additional difficulties in sampling non-decomposable graphical models are
\cite{Jones:05}:
firstly, the normalizing constants in the non-complete prime component marginals do not have closed form, so we need Monte Carlo methods to estimate them. Secondly, these Monte Carlo calculated values have high variance; and finally, when single-edge perturbations are made to the graph, there is no guarantee of significant cancellations in likelihood ratios.

These difficulties hugely increase computing time: in the experiments of \textcite{Jones:05}, 420 times for a 12-node, 15-edge example; 5500 times for 15-node, 26-edge example; this is for Gaussian models, using conjugate priors on variances. These timings are based on their implementation of the Giudici--Green sampler for the decomposable graph problem, so the stated factors would increase by about one order of magnitude if the junction tree samplers were used instead.

\textcite{Jones:05} concluded that sampling from the posterior is not practical for problems with much more than 15 nodes; they resort to fast heuristics like stochastic shotgun search to identify a graph with high posterior probability instead. 

Later results in other recent papers might be used in conjunction with the graph algorithms in \textcite{Jones:05}. \textcite{WangC:10} present a direct method of sampling from the hyper-inverse Wishart distribution on non-decomposable graphs. \textcite{MitsakakisME:11} discuss Metropolis--Hastings sampling of $\Sigma^{-1}$ from a G-Wishart distribution for a fixed $G$. They show how the Deviance Information Criterion can be estimated from such a sample and hence used for model selection. However, the only search scheme considered is complete enumeration, so the approach is again limited to very small graphs; they present an example with 150 observations on 4 variables. 

However, there remains no indication that full joint probabilistic inference about structure and parameters is possible when there are more than, say, 20 vertices; we consider that in consequence there is a continued need for inference restricted to models assuming decomposability.

\section*{Appendices}

\subsection*{Appendix 1: Proofs of decomposability}
Here we provide proofs that the modified graphs $G'$ in Section \ref{sec:multi-edge} are decomposable.

These proofs are constructive but indirect; we actually demonstrate that the described multiple-edge connections and disconnections can be implemented by manipulating a junction tree representing the given decomposable graph; by showing that the result is a valid junction tree we will have shown that the modified graph is decomposable. The precise manipulations to the junction tree are specified algorithmically in Sections \ref{sec:mc} and \ref{sec:md}, and these should be considered in parallel with Propositions 1 and 2 respectively.

It is clear that both the multiple-edge connect and disconnect moves take the current junction tree $J$ and yield a modified graph $J'$ that is still a tree, whose nodes are sets of vertices of $G$. From consideration of the algorithm specification and stated requirements about various sets of vertices being non-empty, it is clear that these nodes of $J'$ are cliques in $G'$. To prove that the corresponding modified graph $G'$ remains decomposable it is therefore sufficient to show that $J'$ still has the junction property, and for this it is sufficient to show that for every vertex $z\in V$, the cliques containing $z$ form a connected sub-tree of $J'$, given that this is true of $J$.

\begin{proof}[of Proposition \ref{Cprop}]
We consider the four cases (a), (b), (c), (d) described in Section \ref{sec:mc} in turn, in each case considering the possibilities that $z$ is in $X$, $Y$, $S$ or $V\setminus (X\cup Y\cup S)$. In case (a), the cliques in $J$ containing $z$ for $z\in X$ are $XS$ and possibly others forming a sub-tree including $XS$; in $J'$, $XS$ is replaced by $XYS$, with the same adjacencies, and this new clique still contains such $z$. For $z\in Y$, the argument is identical; for $z\in S$, the adjacent cliques $XS$ and $YS$ containing $z$ are merged into $XYS$ whose adjacencies combine those of $XS$ and $YS$, so adjacencies among all cliques containing $z$ are preserved. For $z\in V\setminus (X\cup Y\cup S)$, there is no change to the cliques containing $z$ or their adjacencies. In case (b) the only change to $J$ is that vertices in $X$ are added into the clique $YS$, which is adjacent to $XS$ in $J$ so the connected sub-tree property is maintained. Case (c) is similar. Finally, in case (d), the change in $J'$ is that an additional clique $XYS$ is inserted between $XS$ and $YS$: since this is the union of these two cliques, this change cannot affect the connectedness of the sub-trees containing any vertex.
\end{proof}
\begin{proof}[of Proposition \ref{Dprop}]
The arguments about validity of the multiple-edge disconnections proceed along similar lines to those used above. Vertices outside $X\cup Y\cup S$ are not affected by the changes to $J$. In case (a), the requirement to connect cliques in $\N_X$ to $XS$ and those in $\N_Y$ to $YS$, described in Section 3.2(a), ensures connectedness of the sub-trees containing vertices in $X\cup Y$, while those vertices in $S$ are included in all of the new parts of the junction tree. In case (b) and (c) we are removing vertices, in $X$ and $Y$ respectively, from the clique $XYS$; but by assumption $XS$, respectively $YS$, is the only adjacent clique intersecting $X$, respectively $Y$, so all adjacencies are maintained. Finally in case (d), we remove the clique $XYS$ and make $XS$ and $YS$ adjacent. This cannot break any adjacencies.
\end{proof}

\subsection*{Appendix 2: Sampling data from the graphical Gaussian intra-class model}

We can easily draw samples from the distribution $N(0,\Sigma_G(\sigma^2,\rho))$.
For any clique $C$ and separator $S$ such that $S\subset C$,
$$
p(y_{C\setminus S}\mid y_S,G,\sigma^2,\rho) = \frac{p(y_C\mid G,\sigma^2,\rho)}{p(y_S\mid G,\sigma^2,\rho)}
$$
and after using (\ref{eq:ggimcliq}) for both numerator and denominator, and simplifying, we find this can be written
$$
y_{C\setminus S}\mid y_S,G,\sigma^2,\rho \sim N\left(\frac{\rho}{1-\rho+v_S\rho} (\sum_{i\in S} y_i) 1_{C\setminus S},
(1-\rho)\sigma^2(I_{C\setminus S}+\frac{\rho}{1-\rho+v_S\rho}J_{C\setminus S})\right),
$$
where $1_{C\setminus S}$ is a vector of 1's appropriately indexed.

This can be used recursively along the branches of a junction tree to simulate a draw from $N(0,\Sigma_G(\sigma^2,\rho))$.

\subsection*{Appendix 3: Likelihood ratios for the graphical Gaussian intra-class model}

Certain likelihood ratios, ratios of the joint density of $Y$ for two different $G$, can simplify greatly, using the clique-separator factorisation (\ref{eq:cliqsepgauss}). For example, for disjoint sets $A$, $B$ and $S$, writing, e.g. $AS$ for $A\cup S$, 
\begin{multline*}
\frac{p(y_{ABS}\mid G,\sigma^2,\rho)p(y_S\mid G,\sigma^2,\rho)}{p(y_{AS}\mid G,\sigma^2,\rho)p(y_{BS}\mid G,\sigma^2,\rho)}
=\\
\left[\{f(AS)f(BS)\}/\{f(ABS)f(S)\}\right]^{1/2}  \times 
\exp\left(\frac{\rho}{2\sigma^2(1-\rho)}
\left\{H(ABS)+H(S)-H(AS)-H(BS)\right\}
\right).
\end{multline*}
This is the cross-ratio relevant to a single observation. 

Given replicate observations $y^{(r)}\sim N(0,\Sigma_G(\sigma^2,\rho))$, independently for $r=1,\ldots,n$, we need the ratio
\begin{multline*}
\prod_{r=1}^n \left(\frac{p(y_{ABS}^{(r)}\mid G,\sigma^2,\rho)p(y_S^{(r)}\mid G,\sigma^2,\rho)}{p(y_{AS}^{(r)}\mid G,\sigma^2,\rho)p(y_{BS}^{(r)}\mid G,\sigma^2,\rho)}\right)
=\\
\left[\{f(AS)f(BS)\}/\{f(ABS)f(S)\}\right]^{n/2}  \times 
\exp\left(\frac{\rho}{2\sigma^2(1-\rho)}
\left\{H(ABS)+H(S)-H(AS)-H(BS)\right\}
\right),
\end{multline*}
where now $H(D)=\sum_{r=1}^n(\sum_{i\in D} y_i^{(r)})^2/(1-\rho+v_D\rho)$ for each $D$.


\begin{thebibliography}{}

\bibitem[\protect\citeAY{Abel and Thomas}{2011}]{Abel+Thomas:11}
Abel, H.~J. and Thomas, A. (2011).
\newblock Accuracy and computational efficiency of a graphical modeling
  approach to linkage disequilibrium estimation.
\newblock {\em Statistical Applications in Genetics and Molecular Biology},
  {\bf 10}.
\newblock Article 5.

\bibitem[\protect\citeAY{Atay-Kayis and Massam}{2005}]{Atay-Kayis:05}
Atay-Kayis, A. and Massam, H. (2005).
\newblock A {Monte Carlo} method for computing the marginal likelihood in
  nondecomposable graphical {G}aussian models.
\newblock {\em Biometrika}, {\bf 92}, 317--35.

\bibitem[\protect\citeAY{Carvalho and Scott}{2009}]{CarvalhoS:09}
Carvalho, C.~M. and Scott, J. (2009).
\newblock Objective {B}ayesian model selection in {G}aussian graphical models.
\newblock {\em Biometrika}, {\bf 96}, 497--512.

\bibitem[\protect\citeAY{Cayley}{1889}]{Cayley:89}
Cayley, A. (1889).
\newblock A theorem on trees.
\newblock {\em Quarterly Journal of Mathematics}, {\bf 23}, 376--8.

\bibitem[\protect\citeAY{Dellaportas and Forster}{1999}]{Dellaportas:99}
Dellaportas, P. and Forster, J. (1999).
\newblock {Markov chain Monte Carlo} model determination for hierarchical and
  graphical log-linear models.
\newblock {\em Biometrika}, {\bf 86}.

\bibitem[\protect\citeAY{Dellaportas {\it et~al}.}{2003}]{Dellaportas:03}
Dellaportas, P., Giudici, P., and Roberts, G.~O. (2003).
\newblock Bayesian inference for nondecomposable graphical {G}aussian models.
\newblock {\em Sankhy\={a}}, {\bf 65}, 43--55.

\bibitem[\protect\citeAY{Dobra {\it et~al}.}{2011}]{DobraLR:11}
Dobra, A., Lenkoski, A., and Rodriguez, A. (2011).
\newblock Bayesian inference for general {G}aussian graphical models with
  application to multivariate lattice data.
\newblock {\em Journal of the American Statistical Association}, {\bf 106},
  1418--33.

\bibitem[\protect\citeAY{Frydenberg and Lauritzen}{1989}]{Frydenberg+L:89}
Frydenberg, M. and Lauritzen, S.~L. (1989).
\newblock Decomposition of maximum likelihood in mixed interaction models.
\newblock {\em Biometrika}, {\bf 76}, 539--55.

\bibitem[\protect\citeAY{Giudici and Green}{1999}]{Giudici+Green:99}
Giudici, P. and Green, P.~J. (1999).
\newblock Decomposable graphical {Gaussian} model determination.
\newblock {\em Biometrika}, {\bf 86}, 785--801.

\bibitem[\protect\citeAY{Green}{1995}]{Green:95}
Green, P.~J. (1995).
\newblock Reversible jump {Markov chain Monte Carlo} computation and {B}ayesian
  model determination.
\newblock {\em Biometrika}, {\bf 82}, 711--32.

\bibitem[\protect\citeAY{Grone {\it et~al}.}{1984}]{Grone:84}
Grone, R., Johnson, C.~R., S\'a, E.~M., and Wolkowicz, H. (1984).
\newblock Positive definite completions of partial {H}ermitian matrices.
\newblock {\em Linear Algebra and its Applications}, {\bf 58}, 109--24.

\bibitem[\protect\citeAY{Hastings}{1970}]{Hastings:70}
Hastings, W.~K. (1970).
\newblock {Monte} {Carlo} sampling methods using {Markov} chains and their
  applications.
\newblock {\em Biometrika}, {\bf 57}, (1), 97--109.

\bibitem[\protect\citeAY{Jones {\it et~al}.}{2005}]{Jones:05}
Jones, B., Carvalho, C., Dobra, A., Hans, C., Carter, C., and West, M. (2005).
\newblock Experiments in stochastic computation for high dimensional graphical
  models.
\newblock {\em Statistical Science}, {\bf 20}, 388--400.

\bibitem[\protect\citeAY{Lauritzen}{1996}]{Lauritzen:96}
Lauritzen, S.~L. (1996).
\newblock {\em Graphical Models}. Clarendon Press, Oxford.

\bibitem[\protect\citeAY{Lenkoski and Dobra}{2011}]{LenkoskiD:11}
Lenkoski, A. and Dobra, A. (2011).
\newblock Computational aspects related to inference in {G}aussian graphical
  models with the {G-Wishart} prior.
\newblock {\em Journal of Computational and Graphical Statistics}, {\bf 20},
  140--57.

\bibitem[\protect\citeAY{Mitsakakis {\it et~al}.}{2011}]{MitsakakisME:11}
Mitsakakis, N., Massam, H., and Escobar, M.~D. (2011).
\newblock A {Metropolis--Hastings} based method for sampling from the
  {G-Wishart} distribution in {G}aussian graphical models.
\newblock {\em Electronic Journal of Statistics}, {\bf 5}, 18--30.

\bibitem[\protect\citeAY{Peskun}{1973}]{Peskun:73}
Peskun, P.~H. (1973).
\newblock Optimum {Monte-Carlo} sampling using {Markov} chains.
\newblock {\em Biometrika}, {\bf 60}, (3), 607--12.

\bibitem[\protect\citeAY{Rajaratnam {\it et~al}.}{2008}]{RajaratnamMC:08}
Rajaratnam, B., Massam, H., and Carvalho, C.~M. (2008).
\newblock Flexible covariance estimation in graphical {G}aussian models.
\newblock {\em The Annals of Statistics}, {\bf 36}, 2818–--2849.

\bibitem[\protect\citeAY{Roverato}{2002}]{Roverato:02}
Roverato, A. (2002).
\newblock Hyper-inverse {W}ishart distribution for non-decomposable graphs and
  its application to {B}ayesian inference for {G}aussian graphical models.
\newblock {\em Scandinavian Journal of Statistics}, {\bf 29}, 391--411.

\bibitem[\protect\citeAY{Scott and Carvalho}{2008}]{ScottC:08}
Scott, J.~G. and Carvalho, C.~M. (2008).
\newblock Feature-inclusion stochastic search for {G}aussian graphical models.
\newblock {\em Journal of Computational and Graphical Statistics}, {\bf 17},
  790–--808.

\bibitem[\protect\citeAY{Tarantola}{2004}]{Tarantola:04}
Tarantola, C. (2004).
\newblock {MCMC} model determination for discrete graphical models.
\newblock {\em Statistical Modelling}, {\bf 4}, 39--61.

\bibitem[\protect\citeAY{Tarjan and Yannakakis}{1984}]{Tarjan+Yannakakis:84}
Tarjan, R.~E. and Yannakakis, M. (1984).
\newblock Simple linear-time algorithms to test chordality of graphs, test
  acyclicity of hypergraphs, and selectively reduce acyclic hypergraphs.
\newblock {\em SIAM Journal of Computing}, {\bf 13}, 566--79.

\bibitem[\protect\citeAY{Thomas and Green}{2009}]{Thomas+Green:09e}
Thomas, A. and Green, P.~J. (2009).
\newblock Enumerating the junction trees of a decomposable graph.
\newblock {\em Journal of Computational and Graphical Statistics}, {\bf 18},
  930--40.

\bibitem[\protect\citeAY{Wang and Carvalho}{2010}]{WangC:10}
Wang, H. and Carvalho, C.~M. (2010).
\newblock Simulation of hyper-inverse {W}ishart distributions for
  non-decomposable graphs.
\newblock {\em Electronic Journal of Statistics}, {\bf 4}, 1470--5.

\end{thebibliography}

\end{document}